\documentclass[epj]{svjour}
\usepackage{latexsym}
\usepackage{graphics}
\usepackage{color}
\def\Journal#1#2#3#4{ #1 {\bf #2}, #3 (#4)}
\def\eprint#1{arxiv: #1}

\newcommand{\NPB}{Nucl. Phys. B}
\newcommand{\PLB}{Phys. Lett. B}
\newcommand{\PRL}{Phys. Rev. Lett.}
\newcommand{\PRD}{Phys. Rev. D}
\newcommand{\PRP}{Phys. Rep.}

\newcommand{\JHEP}{JHEP}

\newcommand{\IJMPA}{Int. J. Mod. Phys. A}

\newcommand{\CPC}{Comput. Phys. Commun.}
\newcommand{\EPJC}{Eur. Phys. J. C}
\newcommand{\RMP}{Rev. Mod. Phys.}

\journalname{Eur. Phys. J. C}
\begin{document}

\title{Two-parametric observables for Abelian \boldmath $Z'$ searching in $p\bar{p}$ collisions at Tevatron energies
}

\titlerunning{Two-parametric observables for \boldmath $Z'$ searching} 

\author{A.V.~Gulov\inst{1}
        \and
        A.A.~Kozhushko\inst{2} 
}

\institute{Dnipropetrovsk National University, 49010 Dnipropetrovsk, Ukraine, \email{alexey.gulov@gmail.com} \and Dnipropetrovsk National University, 49010 Dnipropetrovsk, Ukraine, \email{andrey.a.kozhushko@gmail.com}}

\date{Received: date / Accepted: date}
\abstract{
We propose a scheme of searches for the Abelian $Z'$ gauge boson in the Drell-Yan process at $\sqrt{S}=1.96$ TeV. We base our considerations on renor\-ma\-li\-za\-tion-group relations between the $Z'$ couplings to standard-model fermions. Considering the range of energies near the $Z$-boson peak, namely 66--116 GeV for the invariant mass of a leptonic pair, we propose an integration scheme to construct two-parametric observables suitable for $Z'$ searches in the $p\bar{p} \to l^+ l^-$ scattering. The observables allow to constrain the $Z'$ vector and axial-vector couplings to SM fermions in a general phenomenological parameterization with non-universal $Z'$ interactions with fermion generations. We also consider the cases of generation-universal $Z'$ boson and leptophobic $Z'$ boson, and show that one-parametric observables exist for these scenarios. The research is aimed to supplement direct searches for $Z'$ as an on-shell state in a specific set of new-physics models.
\PACS{
		{11.10.Gh}{Renormalization} \and
		{12.60.Cn}{Extensions of electroweak gauge sector} \and
		{14.70.Pw}{Other gauge bosons}
	 }
\keywords{Abelian $Z'$ -- Drell-Yan process -- Tevatron}
}

\maketitle

\section{\label{sec:intro}Introduction}

A new heavy neutral vector boson ($Z'$ boson) \cite{leike,Lang08,Rizzo06,LangackerLuo1992,HewettRizzo1988} is a popular scenario of searching for physics beyond the standard model (SM) of elementary particles in modern collider experiments.
Both the Tevatron and LHC collaborations attempt to catch the particle as a resonance in the Drell-Yan process considering some predefined set of $Z'$ models. Observing no peak they conclude that the $Z'$ mass is no less than approximately 2.6 TeV \cite{CMS:2014zpr,ATLAS:2014zpr}. 

Another approach is to search for $Z'$ in processes where it manifests itself as a virtual state. This includes the processes with the so called low-energy neutral currents (LENC) mediated by $Z$ boson. For example, the data on parity violation in cesium can be used to constrain the $Z'$ mass \cite{Derevianko2009} (assuming no mixing between $Z$ and $Z'$). The combined analysis of data on atomic parity violation, inelastic neutrino scattering, and neutrino-electron scattering \cite{Cho1998:1,Cho1998:2,Barger1998} allowed to constrain Fermi-like couplings that effectively represent $Z'$-mediated interactions at low energies \cite{Gulov:1999ry}. General review of low-energy constraints on the $Z'$ boson is presented in \cite{Erler:2009jh} and in section 10 of Ref. \cite{rpp}.

Significant amount of the Tevatron data is collected at the $Z$-boson peak at 66--116 GeV. At these energies the $Z'$ boson manifests itself primarily via $Z-Z'$ mixing, and the ideas to look for signals of extra neutral gauge bosons in this region were expressed and studied earlier \cite{CveticLynnLagr}. Theoretically, these effects should either allow to discover $Z'$ signal, or to constrain its parameters by fitting the experimental data.

In order to select $Z'$ off-shell hints, proper observables have to be introduced to amplify possible signal, e.g. as it was done in \cite{Gulov:1999ry,Pankov98:1,Pankov98:2,Pankov98:3}. The signal generally means a deviation of some $Z'$ parameter (i.e. a coupling) from zero at a specified confidence level. The larger number of such parameters interfere in the observable, the weaker constraints on the parameters are obtained. Thus, the key problem for off-shell $Z'$ detection is to reduce the number of $Z'$ couplings in the observable that is used for data fitting. The ultimate scenario assumes a one-parametric observable. However, a two-parametric observable can be also useful and effective.
For example, the strategy to construct observables driven by one or two parameters was successfully applied to analyze the final data of the LEP experiment leading to hints and constraints on $Z'$ couplings \cite{Gulov:1999ry,Gulov:2004prd,Gulov:2007prd}. So, attempts of selecting possible $Z'$ signals from Tevatron data seem perspective. 

In this paper we construct few-parametric observables for the proton-antiproton scattering processes at $\sqrt{S} = 1.96$~TeV. We work in a framework of Abelian $Z'$. The general case of a $Z'$ boson with non-universal pheno\-menological $Z'$ couplings to fermion generations is considered. When combined with quark mixing effects, the non-universality feature can lead to appearance of the flavor-changing neutral currents, which are strongly suppressed by modern experiments. However, such a discussion requires to specify a complete particle content beyond the SM, which is outside the scope of usual phenomenological parameterizations. In this regard, a non-universal $Z'$ is also a very popular scenario for new physics searches both in high- and low-energy experiments (see references in the survey by P. Langacker \cite{Lang08}) and is motivated by several string models \cite{nonuniv_string:1,nonuniv_string:2,nonuniv_string:3,nonuniv_string:4}. The universality of couplings leads to a different (reduced) initial set of couplings. Such parameterization is also considered. We obtain several possible two-parametric observables at energies corresponding to $Z$ peak. These observables can be used as a key in searches for possible signals of the off-shell $Z'$ boson. Data fitting is a subject of separate investigation and lies beyond the scope of the paper. Let us note that our suggestions and results are valid both for the minimal SM and for the two-Higgs doublet model (THDM).

The paper is organized as follows. In Section~\ref{sec:abelianzpr} we provide all necessary information on the low-energy $Z'$ parameterization for our calculations. Section~\ref{sec:ZprDY} contains specifics on $Z'$ contribution to the Drell-Yan process, uncertainties, and kinematic variables used in had\-ron collider experiments. In Section~\ref{sec:theobservable} we construct the observables in a step-by-step manner. Section \ref{app:universality} is devoted to a special case of generation-universal $Z'$ boson. Section~\ref{sec:discussion} presents a brief discussion the obtained results. In Section~\ref{sec:conclusions} we summarize the proposed integration scheme. Appendix \ref{app:kfactor} contains description of encountered computation difficulties. In Appendix \ref{app:cs} we explicitly write out the partonic cross sections in the used parameterization.



\section{\label{sec:abelianzpr}Abelian $Z'$ couplings to leptons and quarks}

Being decoupled at energies of order of $m_Z$, the Abelian $Z'$ boson interacts with the SM particles as an additional $\tilde{U}(1)$ gauge boson. Its couplings to the SM fermions are usually parameterized by the effective Lagrangian:
\begin{eqnarray}\label{ZZplagr}
{\cal L}_{Z\bar{f}f}&=&\frac{1}{2} Z_\alpha\bar{f}\gamma^\alpha\left[
(v^\mathrm{SM}_{fZ}+\gamma^5 a^\mathrm{SM}_{fZ})\cos\theta_0 
\right. \nonumber\\
&&\left. 
+(v_f+\gamma^5 a_f)\sin\theta_0 \right]f, \nonumber\\
{\cal L}_{Z'\bar{f}f}&=&\frac{1}{2} Z'_\alpha\bar{f}\gamma^\alpha\left[
(v_f+\gamma^5 a_f)\cos\theta_0 
\right. \nonumber\\
&&\left. 
-(v^\mathrm{SM}_{fZ}+\gamma^5
a^\mathrm{SM}_{fZ})\sin\theta_0\right]f.
\end{eqnarray}
Here $f$ is an arbitrary SM fermion state;
$a_f$ and $v_f$ are the $Z'$ couplings to the
axial-vector and vector fermion currents, respectively; 
$v^\mathrm{SM}_{fZ}$,
$a^\mathrm{SM}_{fZ}$ are the SM couplings of the $Z$ boson; 
$\theta_0$ is the $Z$--$Z'$ mixing angle. 
The $a_f$ and $v_f$ couplings are proportional to the $Z'$ gauge coupling $\tilde{g}$. This Lagrangian is inspired by adding new 
$\tilde{U}(1)$ terms to the common covariant derivatives 
$D^\mathrm{EW}$ used in the SM \cite{CveticLynnLagr,DegrassiSirlinLagr}.
If the minimal standard model is considered as the low-energy theory, the mixing angle
relates to the generator corresponding to the $\tilde{U}(1)$ gauge group, 
$\tilde{Y}_{\phi}$, as
\begin{equation}\label{eq:theta1}
\theta_0 =
\frac{\tilde{g}\sin\theta_W\cos\theta_W}{\sqrt{4\pi\alpha_\mathrm{em}}}
\frac{m^2_Z}{m^2_{Z'}} \tilde{Y}_\phi
+O\left(\frac{m^4_Z}{m^4_{Z'}}\right),
\end{equation}
Eq. (\ref{eq:theta1}) originates from diagonalization of the mass matrix of neutral gauge bosons. 
In case of the Abelian $Z'$ boson this result is also valid for the THDM \cite{Gulov:2000epjc}.

This parameterization is suggested by a number of natural conditions:
(i) The $Z'$ interactions of renormalizable types are to
be dominant at low energies $\sim m_Z$. The non-renormalizable
interactions generated at high energies due to radiation
corrections are suppressed by the inverse heavy mass
$1/m_{Z^\prime}$ (or by other heavier scales $1/\Lambda_i\ll
1/m_{Z^\prime}$). Therefore, one can neglect such interactions.
(ii) The $Z^\prime$ boson is the only neutral vector
boson with the mass $\sim m_{Z^\prime}$.

Below the $Z'$ decoupling threshold the effective $\tilde{U}(1)$ symmetry is a trace of the renormalizability of an unknown complete model with the $Z'$ boson, and it leads to additional relations between the $Z'$ couplings 
\cite{Gulov:2000epjc,GulovSkalozub:2009review,GulovSkalozub:2010ijmpa}:
\begin{equation} \label{grgav}
v_f - a_f= v_{f^*} - a_{f^*}, \qquad a_f = T_{3f}
\tilde{g}\tilde{Y}_\phi,
\end{equation}
where $f$ and $f^*$ are the partners of the $SU(2)_L$ fermion
doublet ($l^* = \nu_l, \nu^* = l, q^*_u = q_d$ and $q^*_d = q_u$);
$T_{3f}$ is the third component of weak isospin; 
$\tilde{g}\tilde{Y}_\phi$ includes the $Z'$ gauge coupling and determines 
the $Z'$ interactions to the SM scalar fields.

The relations (\ref{grgav}) are conveniently rewritten as
\begin{eqnarray}\label{RGrel1}
&&
a_{q_d} = a_{l} = -a_{q_u} = -a_{\nu_l}=a,\nonumber\\
&&
v_{q_d} = v_{q_u} + 2 a, \qquad v_{l} = v_{\nu_l} + 2 a,
\end{eqnarray}
where $q_u$, $q_d$, $l$, and $\nu_l$  are an up-type and a down-type quark, a lepton, and a neutrino within a fermion generation, respectively. Coupling $a$ is a universal coupling constant that defines also 
the $Z$--$Z'$ mixing angle in (\ref{ZZplagr}). By substituting Eqs. (\ref{grgav}) into (\ref{eq:theta1}) we obtain
\begin{equation}\label{RGrel2}
\theta_0 \approx -2a\frac{\sin \theta_W \cos
\theta_W}{\sqrt{4 \pi \alpha_{\rm em}}} \frac{m_Z^2}{m_{Z'}^2}.
\end{equation}
The discussed relations (\ref{grgav})--(\ref{RGrel2}) are also true 
for the THDM case. More details on this matter can be found in \cite{Gulov:2000epjc}. The full Lagrangian is written out in
\cite{GulovKozhushko:2011ijmpa}.

As it was shown in
\cite{GulovSkalozub:2009review,GulovSkalozub:2010ijmpa}, the
relations (\ref{grgav}), (\ref{RGrel1}) cover, in particular, a bunch of GUT models based on
the ${E}_6$ group (the $\chi$ model) and the $SO(10)$ group (the so called {\it left-right} model).
It has to be noted that these relations are not obligatory for any $Z'$ model 
discussed in the literature. Nevertheless, they
describe correlations between $Z'$ couplings for a wide set of
models beyond the SM.
Same applies to the mixing angle. It is easy to check that the relation (\ref{RGrel2}) stands for the $\chi$ model and the left-right model by substituting axial couplings  provided e.g. in \cite{leike,Lang08,Rizzo06,LangackerLuo1992}.

As a result, Abelian $Z'$ couplings can be parameterized by seven
independent parameters $a$, $v_u$, $v_c$,
$v_t$, $v_e$, $v_\mu$, $v_\tau$.
These parameters must be fitted in
experiments. In a particular model, one has some specific values
for them. In case when the model is unknown, these parameters
remain potentially arbitrary numbers. 

\section{\label{sec:ZprDY}Abelian $Z'$ in the Drell-Yan process}

At the Tevatron the most prominent signal of the Abelian $Z'$ boson is expected in the $p\bar{p} \to l^+ l^-$ scattering process. The general idea of our approach is equally applicable both for dielectrons and dimuons in the final state, therefore we will discuss lepton pairs in general.
The cross section of this process is commonly written in form of the partonic cross sections combined with the parton distribution functions (PDFs):
\begin{eqnarray}
\frac{\partial^3 \sigma_{AB}}{\partial x_q \partial x_{\bar{q}} 
\partial \hat{t}} &=&
\sum_{q,\bar{q}}\, f_{q,A}
(x_q,\mu_\mathrm{F},\mu_\mathrm{R})f_{\bar{q},B}(x_{\bar{q}},\mu_\mathrm{F},\mu_\mathrm{R}) 
\nonumber\\
&\times &
\frac{\partial \hat{\sigma}_{q\bar{q}\to l^+ l^-}}{\partial \hat{t}},\nonumber\\
\hat{\sigma}_{q\bar{q}\to l^+ l^-} &=& \hat{\sigma}_{q\bar{q}\to l^+ l^-}(\hat{t}),
\end{eqnarray}
where $A$, $B$ mark the interacting hadrons ($p$ or $\bar{p}$)
with the four-momenta $k_A$, $k_B$; multipliers $f_{q,A}(x_q, \mu_\mathrm{F},\mu_\mathrm{R})$ are the PDFs for the parton $q$ in the hadron $A$ with
the momentum fraction $x_q$ ($0 \leq x_q \leq 1$) at the factorization
scale $\mu_\mathrm{F}$ and renormalization scale $\mu_\mathrm{R}$. To access the parton distribution data, we use the MSTW 2008 package \cite{mstw:1,mstw:2}. 
The quantity $\hat{\sigma}_{q\bar{q}\to l^+ l^-}$ is the parton-level 
cross section, which depends on the Mandelstam variable $\hat{t} = 
(p_{l^+} - p_{q})^2$. 
All parton-level calculations are performed using \textit{FeynArts} \cite{FeynArts:1,FeynArts:2} and \textit{FormCalc} \cite{FormCalc:1,FormCalc:2} packages. Hereafter, the hat over a variable denotes that this variable refers to the parton-level cross section.

Let us denote the PDF factor for each quark flavor as:
\begin{eqnarray}
f_{q,A}(x_q,\mu_\mathrm{F,R})f_{\bar{q},B}(x_{\bar{q}},\mu_\mathrm{F,R}) = F_{q\bar{q}}(x_q,x_{\bar{q}},\mu_\mathrm{F,R}).
\end{eqnarray}

The obtained triple-differential cross section provides full description for the Drell-Yan process. It is expressed in terms of three kinematic variables: $x_q$, $x_{\bar{q}}$, and $\hat{t}$. The shortcoming of these variables is that all three of them enter both the PDF multiplier and the parton-level cross section, since the Mandelstam variable $\hat{s} = (p_{l^+} + p_{l^-})^2$ is not an independent value ($\hat{s} = x_q x_{\bar{q}} S$).

The quantities that are directly measured in experiments and used for event selection are the pseudorapidities $\eta_{\pm}$ and transverse momenta $p_{T}^{\pm}$ of the final-state leptons. In the leading order in $\alpha_S$ the relation $p_{T}^+ = -p_{T}^- = p_T$ applies.
The Mandelstam variables $\hat{s}$, $\hat{t}$ and the momentum fractions $x_q$, $x_{\bar{q}}$ are expressed as
\begin{eqnarray}
&& \hat{s} = M^2 = 4 p_T^2 \, \cosh^2 \, \frac{\eta_{+}-\eta_{-}}{2}, \quad
\hat{t} = -\frac{M^2}{1+ e^{(\eta_{+}-\eta_{-})}}, \nonumber\\
&& x_q = \frac{M}{\sqrt{S}} e^{(\eta_{+}+\eta_{-})/2}, \quad
x_{\bar{q}} = \frac{M}{\sqrt{S}} e^{-(\eta_{+}+\eta_{-})/2}.
\end{eqnarray}
Since $x_q$ and $x_{\bar{q}}$ depend only on the sum of the lepton pseudorapidities, while $\hat{t}$ is expressed in terms of the difference of the pseudorapidities, we make a well-known substitution:
\begin{eqnarray}
Y = (\eta_{+}+\eta_{-})/2, \quad
y = (\eta_{+}-\eta_{-})/2. \nonumber
\end{eqnarray}

The $Y$ variable is the intermediate-state rapidity, while $y$ is related to the scattering angle in the $q\bar{q} \to l^+ l^-$ process as 
\begin{eqnarray}
\cos\hat{\theta} = \tanh{y} \nonumber
\end{eqnarray}
and governs the parton-level kinematics (it is introduced in some textbooks, for example in \cite{PeskinSchroeder}). In this way the cross section is obtained as a function of $M$, $Y$, $y$:
\begin{eqnarray}
\label{eq:MYy_cs}
\frac{\partial^3 \sigma_{AB}}{\partial M \partial Y 
\partial y} &=&
\sum_{q,\bar{q}}\, F_{q\bar{q}}(M,Y,\mu_\mathrm{F,R})
\frac{\partial \hat{\sigma}_{q\bar{q}\to l^+ l^-}}{\partial y},\nonumber\\
\hat{\sigma}_{q\bar{q}\to l^+ l^-} &=& \hat{\sigma}_{q\bar{q}\to l^+ l^-}(M,y).
\end{eqnarray}

The explicit expressions for $\hat{\sigma}_{q\bar{q}\to l^+ l^-}$ are quite cumbersome, but it is useful to provide them taking into account the relations (\ref{RGrel1})--(\ref{RGrel2}).
Each $\hat{\sigma}_{q\bar{q}\to l^+ l^-}$ can be written as follows to the $O(\tilde{g}^2)$ order:
\begin{eqnarray}
\label{eq:partcs}
\hat{\sigma}_{q\bar{q}\to l^+ l^-} &=& \hat{\sigma}_\mathrm{SM} + a^2 \hat{\sigma}_{a^2} + a v_l \hat{\sigma}_{a v_l}
\nonumber\\
&&+ a v_u \hat{\sigma}_{a v_u} + 
v_u v_l \hat{\sigma}_{v_u v_l}.
\end{eqnarray}
The $\hat{\sigma}_\mathrm{SM}$ quantity and the factors $\hat{\sigma}_{a^2,a v_l,a v_u,v_u v_l}$ are provided in Appendix \ref{app:cs}. They also can be found in \cite{factorsandmodelfile} along with the model file for the \textit{FeynArts} package. The factors are provided in files suitable for usage in computational packages.

At energies close to $Z$ peak, the leading $Z'$ contribution to the Drell-Yan process arises from mixing between $Z$ and $Z'$ intermediate states, resulting in corrections of order of $O(\tilde{g}^2)$. The contribution of quartic couplings is the $Z'$ on-shell production. At energies much lower than $m_{Z'}$ it is neglected. The cross section reads as
\begin{eqnarray}
\label{eq:zpr_factors}
\sigma_\mathrm{DY} &=& \sigma_\mathrm{SM} + \sigma_{Z'}, \nonumber\\
\sigma_{Z'} &=& a^2 \sigma_{a^2} + a v_l \sigma_{a v_l} 
+ a v_u \sigma_{a v_u} + 
v_u v_l \sigma_{v_u v_l} 
\nonumber\\
&& 
+ a v_c \sigma_{a v_c} + 
v_c v_l \sigma_{v_c v_l}.
\end{eqnarray}
Here $a$, $v_f$ are the couplings defined in (\ref{grgav}) and (\ref{RGrel1}); $\sigma_\mathrm{SM}$ is the standard model contribution to the Drell-Yan process; $\sigma_{a^2}$, $\sigma_{a v_f}$, $\sigma_{v_f v_{f'}}$ are the numerical factors that depend on $M$, $Y$, $y$. In this approximation there are six independent unknown quantities entering the Drell-Yan cross section.
In (\ref{eq:zpr_factors}) the factors that include $v_u$ and $v_c$ arise only from the contributions of first and second generation fermions, respectively. The contribution from the third generation is neglected due to the nature of (anti)protons.

Both the PDF factor and the parton-level cross section are calculated
in the leading order (LO) in $\alpha_S$, and $\sigma_{AB}$ in the LO is obtained in this way. The next-to-next-to-leading
order (NNLO) QCD corrections are then taken into account by multiplying
$\sigma_{AB}$ by the NNLO K-factor, which is calculated using the \textit{FEWZ 2.1.1} software \cite{Petriello:FEWZ} (see also ref. \cite{Petriello:2011FEWZ,Petriello:2012FEWZ}). Since the calculation of the K-factor requires significant computational time, we generate it with some numerical uncertainty and fit the obtained MC data with a polynomial. Please refer to details in Appendix~\ref{app:kfactor}. It is obvious that the K-factor somewhat modifies the $M$- and $Y$-dependence of the LO cross section. At the same time, since we are dealing with dileptonic final state, the internal kinematics of the parton-level process remain unchanged, i.e. the $y$-dependence is the same in LO and in NNLO. This is because the only NNLO corrections in this case arise from virtual-gluon exchanges between the two quarks in the initial state (which affects the $q\bar{q}\gamma^*/Z/Z'$ vertex) and emissions of on-shell gluons by those quarks (which affects the $M$- and $Y$-dependence).
To take into account the electroweak radiative corrections, we introduce decay widths and use the running value of the QED coupling constant $\alpha_{\mathrm{QED}}$ at the $Z$-peak \cite{rpp} of $1/127.9$.

We also consider two kinds of uncertainties: 
(i) The PDF uncertainties $\Delta\sigma_{\mathrm{PDF}}$. The MSTW 2008 package provides
68\% CL and 90\% CL intervals. We consider the latter ones.
(ii) The uncertainties due to the factorization and renormalization scales variation, $\Delta\sigma_{\mu}$. To incorporate these uncertainties, we follow a common procedure: we set $\mu_\mathrm{R} = \mu_\mathrm{F} = \mu$ and vary $\mu$ from $\sqrt{\hat{s}}/2$ to $2\sqrt{\hat{s}}$.

The cross section then can be written as 
\begin{eqnarray}
\sigma_{\mathrm{DY}} = \sigma_{\mathrm{DY}}^{\mathrm{mean}} \pm~\Delta\sigma_{\mathrm{PDF}}\pm\Delta\sigma_{\mu}.
\end{eqnarray}

Once again, we note that $Y$ enters the PDF factors only, while $y$ is included into the parton-level cross sections only. This allows us to treat the PDF factor $F_{q\bar{q}}$ and the partonic cross section $\hat{\sigma}_{q\bar{q}\to l^+ l^-}$ separately.
Therefore, we can try to use any peculiarities in the $M$- and $Y$-dependence of the PDFs and $M$- and $y$-dependence of the partonic cross sections to suppress some of the numerical factors in (\ref{eq:zpr_factors}). For example, if after integration by one of the kinematic variables over some specific range the factor $\sigma_{v_c v_l}$ appears to be much smaller than the other factors, we may neglect its contribution to the cross section and deal with five unknown parameters instead of six we had initially. Of course, we assume that all the combinations of the $Z'$ couplings in the cross section are of the same order of magnitude. The leptophobic $Z'$ case, which is a popular parameterization nowadays, is treated separately in Section \ref{sec:discussion}.

In addition to the $Z'$ couplings, another two unknown $Z'$ parameters affect $\sigma_\mathrm{DY}$. Those are the $Z'$ mass $m_{Z'}$ and decay width $\Gamma_{Z'}$. The latest data from CMS and ATLAS indicates that $Z'$ is heavier than 2.6 TeV. This means that for energies close to the $Z$ peak the $\sigma_\mathrm{DY}$ dependencies on $m_{Z'}$ and $\Gamma_{Z'}$ can be neglected, assuming that the $Z'$ peak is far away from this range. 

The $Y$ and $y$ values that we can investigate are limited by detector performance and conservation laws.
From the condition $0 \leq x_{q,\bar{q}} \leq 1$ it is easy to obtain the $M$-dependent limits
\begin{eqnarray}
\label{eq:conslim}
-\ln \frac{\sqrt{S}}{M} \leq Y \leq \ln \frac{\sqrt{S}}{M}.
\end{eqnarray}

The selection criterion for muons at the D0 Collaboration states that muon pseudorapidity must be in the range $|\eta_\pm| \leq 2.0$ \cite{D0_rapidity_mumu_2011}. Usually, the considered range for electrons and positrons is wider, so we will consider that relatively narrow range. Hence,
\begin{eqnarray}
\label{eq:detlim}
|Y|\leq 2.0.
\end{eqnarray}
The limits for $y$ are the same as for $Y$.

This section can be briefly summarized by saying the following: the cross section of the Drell-Yan process contains six unknown independent terms inspired by $Z'$ boson. The cross section depends on three kinematic variables, which will be used to suppress some of the contributions from the unknown $Z'$ parameters. This will allow us to amplify the signal of $Z'$ that is possibly hidden in the experimental data collected by Tevatron.

\section{\label{sec:theobservable}The Observable}

The most detailed description of a scattering process is contained in the differential cross section. But a possible $Z'$ signal can be washed out because of interference between the six independent combinations of $Z'$ couplings that enter the cross section. In general, integration by kinematic variables can leave this situation unchanged. We need to pay special attention to the integration scheme so as to reduce the number of interfering parameters and to make a successful data fit possible. This scheme must derive benefits from kinematic properties of the cross section.

\subsection{\label{subsec:PDF}Integrating by $Y$}

\begin{figure*}[t]
\centering{
\resizebox{0.32\linewidth}{!}{
\includegraphics{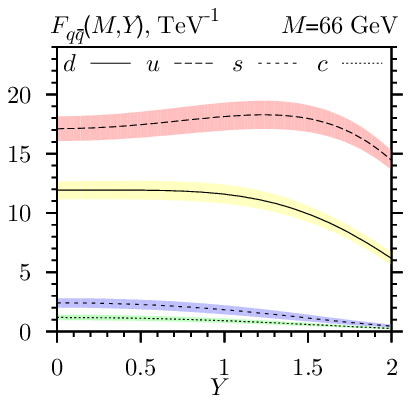}
}
\resizebox{0.32\linewidth}{!}{
\includegraphics{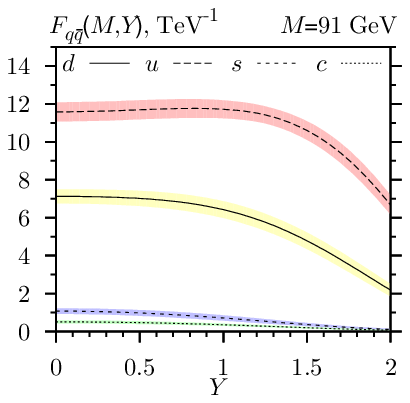}
}
\resizebox{0.32\linewidth}{!}{
\includegraphics{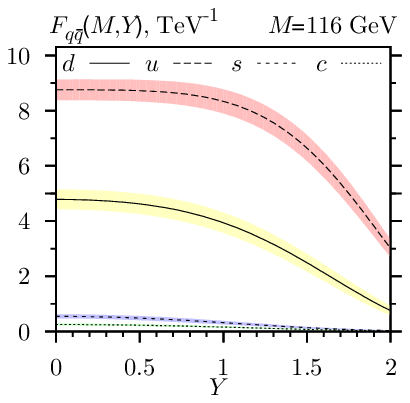}
}
}
\caption{\label{fig:PDFsY}Plots for $F_{q\bar{q}} (M,Y)$ versus $Y$ at different $M$ values. The uncertainties that arise from the PDF errors and factorization scale variation are also shown}
\end{figure*}


The intermediate state rapidity $Y$ enters the PDF factors only. Let us study the $M$- and $Y$-dependence of $F_{q\bar{q}} (M,Y)$ in Eq. (\ref{eq:MYy_cs}). At any fixed kinematically allowed $Y$ value $F_{q\bar{q}}$ is a smooth monotonically decreasing function of $M$.
Kinematic properties of $F_{q\bar{q}}$ are different for each flavor but independent of $Z'$ properties. 
So, the $Y$-dependence of the cross section can be utilized to suppress the contributions of the second generation, i.e. the terms with $a v_c$ and $v_c v_l$ in Eq. (\ref{eq:zpr_factors}). 

We use the following integration scheme
\begin{eqnarray}
\label{eq:observ_Y}
\sigma^{(Y)} = \int_{-Y_\mathrm{max}}^{Y_\mathrm{max}} dY \, W(M,Y)\frac{\partial^3 \sigma_{\mathrm{DY}}}{\partial Y \partial M \partial y}
\end{eqnarray}
 with a simple piecewise-constant weight function
\begin{eqnarray}
\label{eq:observ_Y_wf}
W(M,Y) = \left\{\begin{array}{ll}
W(M),& 0<|Y|\le Y_\mathrm{mid},\\ 
1,& Y_\mathrm{mid}<|Y|<Y_\mathrm{max}.\end{array}\right.
\end{eqnarray}
Here $W(M)$ is a weight function that is manually adjusted for each value of $M$; $Y_\mathrm{mid}$ and $Y_\mathrm{max}$ are some positive integration limits which are also set
manually.

In fact, we integrate the PDF factor in Eq. (\ref{eq:MYy_cs}):
\begin{eqnarray}
\label{eq:PDF_int_Y}
F_{q\bar{q}} (M) &=& 2 \int_0^{Y_m} \,  dY \, W(M,Y) \,
 F_{q\bar{q}} (M,Y),\\
\sigma^{(Y)} &=& \sum_{q,\bar{q}} F_{q\bar{q}}(M)\frac{\partial \hat{\sigma}_{q\bar{q}\to l^+ l^-}}{\partial y}.
\end{eqnarray}
In this subsection we show how to pick the weight function $W(M,Y)$, so that the resulting PDF factors $F_{s\bar{s}} (M)$ and $F_{c\bar{c}} (M)$ are negligibly small compared to $F_{u\bar{u}} (M)$ and $F_{d\bar{d}} (M)$.
Note, that the $Y$-distribution for the Drell-Yan cross section is symmetric. 

Consider $M$ values at the $Z$-peak. Both CDF and D0 collaborations define limits of this range to be symmetric with respect to the $Z$ boson mass. These limits are often set to either $66 \mathrm{~GeV} \leq M \leq 116 \mathrm{~GeV}$ or $71 \mathrm{~GeV} \leq M \leq 111 \mathrm{~GeV}$ \cite{CDF66,D0_zpr_y}. In this paper we use the former option. Actually, the choice of specific lower and upper limits does not affect our results. There are only two general requirements: the limits have to be symmetric with respect to $m_Z$ and large enough so that all quark masses could be set to zero.

In Figure \ref{fig:PDFsY} the plots for $F_{q\bar{q}} (M,Y)$ versus $Y$ at different $M$ values are shown for $u$, $d$, $c$, and $s$ quarks. The relative contributions of second generation quarks amount up to 11\% at $M = 66 \mathrm{~GeV}$ and cannot be neglected. 
For any given $M$ value from the $Z$-peak range we can pick the weight function $W(M,Y)$ in such a way that the factors $F_{c\bar{c}} (M)$ and $F_{s\bar{s}} (M)$ amount to less than 1\% of each of the factors $F_{u\bar{u}} (M)$ and $F_{d\bar{d}} (M)$:
\begin{eqnarray}
\label{eq:pdfs_suppression_condition}
F_{c\bar{c},\,s\bar{s}} (M) \leq 0.01 F_{u\bar{u},\,d\bar{d}} (M)
\end{eqnarray}

In Eq. (\ref{eq:observ_Y_wf}) we set $Y_\mathrm{max} = 2.0$ for consistency with detector 
limitations. We have certain freedom in choosing either $Y_\mathrm{mid}$, or $W(M)$.
We set $Y_\mathrm{mid} = 0.75$ and use the condition (\ref{eq:pdfs_suppression_condition})
to determine $W(M)$ for several $M$ values in the range $66 \mathrm{~GeV} \leq M \leq 116 \mathrm{~GeV}$. After fitting we obtain the following expression for $W(M)$:
\begin{equation}
\label{eq:avsm}
W(M) = 0.547 (M/m_Z) -1.326.
\end{equation}
Here $m_Z$ is the $Z$ mass set to $91.1876$ GeV.

\begin{figure}[b]
\centering{
\resizebox{1\linewidth}{!}{
\includegraphics{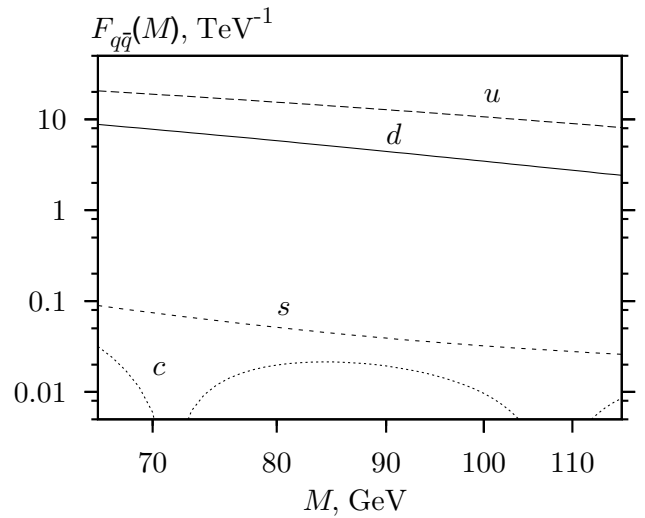}
}}
\caption{\label{fig:ZpeakIntegrated}The plot illustrates suppression of the contributions of second-generation quarks to the Drell-Yan cross section in the $Z$-peak range. The plotted values are $F_{u\bar{u}}$, $F_{d\bar{d}}$, $F_{c\bar{c}}$, and $F_{s\bar{s}}$ integrated by $Y$. The integration by $Y$ is carried out over the range $|Y| \leq 2.0$, where $Y_\mathrm{mid}$ is set to 0.75, and $W(M)$ is determined by Eq. (\ref{eq:avsm})}
\end{figure}

The PDF factors obtained after substituting this weight function into Eq. (\ref{eq:PDF_int_Y}) are shown in Figure \ref{fig:ZpeakIntegrated}. The second-generation contributions, $\sigma_{a v_c}$ and $\sigma_{v_c v_{l}}$, are suppressed. In principle, the values of the weight function parameters are arbitrary. One could set any $Y_\mathrm{max}$ and $Y_\mathrm{mid}$ based on detector performance. The function $W(M)$ is determined based on that choice.

As a result, we obtain the cross section $\sigma^{(Y)}$, which depends on $y$ and $M$ and contains four independent $Z'$ terms instead of six:
\begin{eqnarray}
\label{eq:sigma1}
\sigma^{(Y)} &=& \sigma^{(Y)}_{\mathrm{SM}} + a^2 \sigma^{(Y)}_{a^2} + a v_l \sigma^{(Y)}_{a v_l} 
\nonumber\\
&+& a v_u \sigma^{(Y)}_{a v_u} + 
v_u v_l \sigma^{(Y)}_{v_u v_l}.
\end{eqnarray}

Our next step is to use the two remaining kinematic variables, $M$ and $y$, to get rid of another two unknown combinations of the $Z'$ couplings.

\subsection{\label{subsec:partonic}Integrating by $M$ and $y$}

The variable $y$ enters the parton-level cross section of the Drell-Yan process, $\sigma_{q\bar{q}\to l^+ l^-}$, and is irrelevant for the PDF analysis. In general, the parton-level cross section depends also on $M$ through four `resonant' functions:

\begin{eqnarray}
\label{eq:RF}
f_1(M) &=& \frac{1}{(M^2/m_Z^2 - 1)^2 + \Gamma_Z^2 /m_Z^2}, \nonumber\\
f_2(M) &=& \frac{(M^2/m_Z^2 - 1)}{(M^2/m_Z^2 - 1)^2 + \Gamma_Z^2 /m_Z^2}, \nonumber\\
f_2'(M) &=& \frac{(M^2/m_{Z'}^2 - 1)}{(M^2/m_{Z'}^2 - 1)^2 +
\Gamma_{Z'}^2/m_{Z'}^2},
\nonumber\\
f_3(M) &=& \frac{M^2}{m_Z^2} f_2 f_2' \Bigg(\frac{\Gamma_Z \Gamma_{Z'}/(m_Z m_{Z'})}{(M^2/m_{Z'}^2 - 1)(M^2/m_{Z}^2 - 1)}  \nonumber\\
&& + 1 \Bigg)
\end{eqnarray}

Here $m_{Z, Z'}$ and $\Gamma_{Z, Z'}$ denote the masses and the widths of the $Z$ and $Z'$ bosons. We investigate the energy range close to the $Z$ boson peak. As it was noted earlier, in this case we do not care about specific values of the $Z'$ mass and decay widths. But at this point for numerical calculations we are going to set specific values for $m_{Z'}$ and $\Gamma_{Z'}$. Following the recent LHC results \cite{CMS:2014zpr,ATLAS:2014zpr}, we set $m_{Z'}$ to 2.6 TeV and assume the decay width to be 10\% of the mass. This means that we use some asymptotics of $f'_2$ and $f_3$ at $M\ll m_{Z'}$.

\begin{figure}[t]
\centering{
\resizebox{1\linewidth}{!}{
\includegraphics{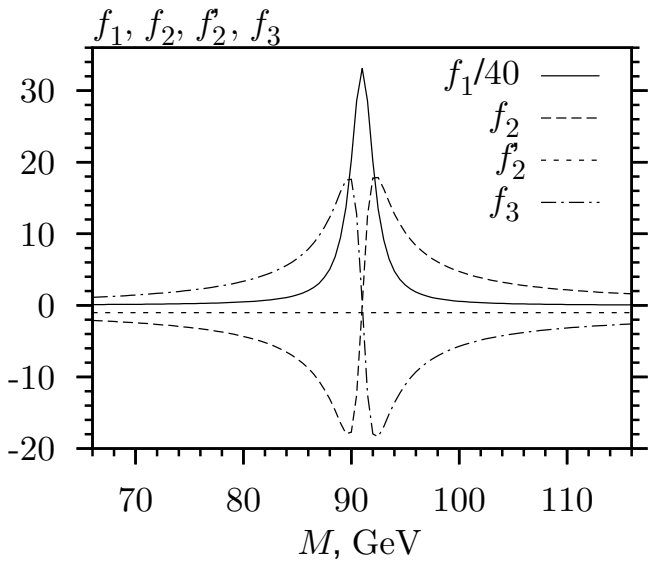}}}
\caption{\label{fig:RF}Plots for the resonant functions given by Eqs. \ref{eq:RF} in the range $66 \mathrm{~GeV} \leq M \leq 116 \mathrm{~GeV}$}
\end{figure}


As it can be seen from Figure \ref{fig:RF}, the $f_1$ function is dominant. 
In the discussed symmetric $Z$-peak range the functions $f_2$, $f_3$ are odd-like with respect to $M=m_Z$, and the function $f'_2$ is small.
As a consequence, after integrating by $M$ over the range the functions $f_2$, $f'_2$, and $f_3$ are negligible compared to $f_1$. Partially this follows from the fact that at the $Z$ peak the leading new physics contribution comes from $Z-Z'$ mixing.
We are going to use the discussed feature in what follows.

\begin{figure*}[t]
\centering{
\resizebox{0.49\linewidth}{!}{
\includegraphics{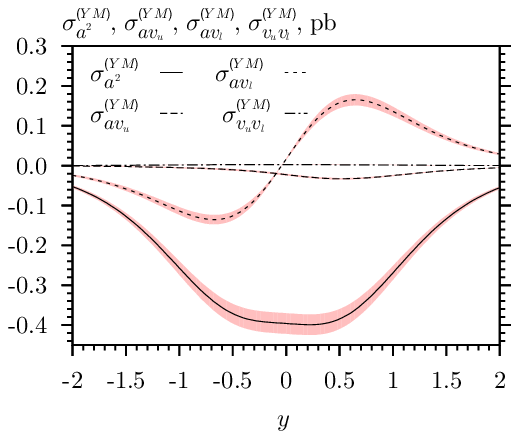}}
\resizebox{0.49\linewidth}{!}{
\includegraphics{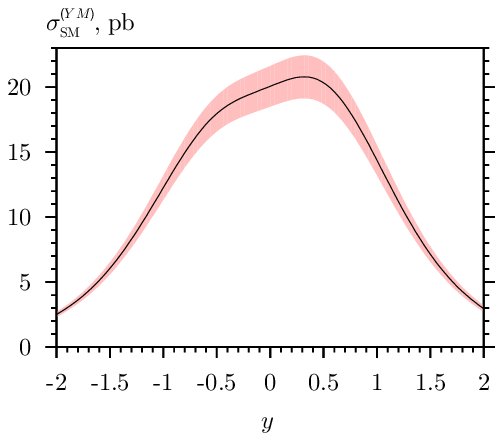}}\\
\hfill (a) \hfill\hfill (b) \hfill}
\caption{\label{fig:factors}Plots for (a) the $Z'$-related factors and (b) $\sigma^{(YM)}_\mathrm{\,SM}$ from Eq. (\ref{eq:sigma2}). The uncertainty bands are also shown}
\end{figure*}


When investigating the $M$-dependence of the hadronic cross section $\sigma^{(Y)}$, we deal not with the resonant functions themselves, but with their products with the PDF factors.
The general form of $\sigma^{(Y)}$ can be written as
\begin{eqnarray}
\label{eq:sigma1_yM}
\sigma^{(Y)} - \sigma^{(Y)}_{\mathrm{SM}}= \frac{\cosh 2y}{\cosh^4 y} \left[ A(M) \tanh 2y + S(M) \right],
\end{eqnarray}
where $A(M)$ and $S(M)$ are some functions that include the unknown couplings $a$, $v_u$, and $v_l$.
The $M$-dependence arises from the `resonant' functions multiplied by $F_{q\bar{q}}(M)$ from Eq. (\ref{eq:PDF_int_Y}). From the plots in Figure~\ref{fig:ZpeakIntegrated} we can conclude that the factors $F_{q\bar{q}}(M)$ are smooth, monotonic, and slowly-varying in the considered mass range. 
Therefore, we stress that all the discussed properties of $f_1$, $f_2$, $f'_2$, and $f_3$ are generally maintained, when these functions are multiplied by $F_{q\bar{q}}(M)$.

Naturally, $f'_2$ and $f_3$ do not enter the SM part $\sigma^{(Y)}_{\mathrm{SM}}$. There are four factors entering the $Z'$ contribution: $\sigma^{(Y)}_{a^2}$, $\sigma^{(Y)}_{a v_l}$, $\sigma^{(Y)}_{a v_u}$, and $\sigma^{(Y)}_{v_u v_l}$ (see Eq. (\ref{eq:sigma1})). The factor $\sigma^{(Y)}_{v_u v_l}$ does not depend on $f_1$. According to our discussion of properties of the `resonant' functions it can be eliminated by straightforward integration by $M$ over the $Z$-peak range (66 GeV $\leq M \leq$ 116 GeV). The resulting value is denoted $\sigma^{(YM)}$:
\begin{eqnarray}
\label{eq:sigma2}
\sigma^{(YM)} - \sigma^{(YM)}_{\mathrm{SM}} & = & \int dM \, (\sigma^{(Y)} - \sigma^{(Y)}_\mathrm{SM}) 
\nonumber\\ 
&=& 
\frac{\cosh 2y}{\cosh^4 y} \left( A \tanh 2y  + S \right), \nonumber\\
\sigma^{(YM)} & = & \sigma^{(YM)}_{\mathrm{SM}} + a^2 \sigma^{(YM)}_{a^2} \nonumber\\
&+& a v_l \sigma^{(YM)}_{a v_l} + a v_u \sigma^{(YM)}_{a v_u},
\nonumber\\
A &=& \int dM \, A(M), \, S = \int dM \, S(M).
\end{eqnarray}
The factors $\sigma^{(YM)}_{\mathrm{SM}}$, $\sigma^{(YM)}_{a^2}$, $\sigma^{(YM)}_{a v_l}$, $\sigma^{(YM)}_{a v_u}$, and $\sigma^{(YM)}_{v_u v_l}$ are plotted on Figure \ref{fig:factors}. It can be seen that $\sigma^{(YM)}_{v_u v_l}$ is negligibly small compared to the other three factors. 

We are not concerned about $\sigma^{(YM)}_{\mathrm{SM}}$ at the moment and shall turn to investigating the $y$-dependence of the $Z'$-related contribution in Eq.~(\ref{eq:sigma2}).
The behavior of the $\sigma^{(YM)}_{a v_u}$ factor is governed by its odd part, while $\sigma^{(YM)}_{a^2}$ is obviously dominated by it's even part.
From the plots on Figure \ref{fig:factors} (a), one can conclude that it is possible to suppress one of the three factors by integrating the cross section by $y$ over a symmetric range. The integration limits for $y$ are the same as for $Y$. In our case 
\begin{equation}
\label{eq:yregion}
-2.0 \leq y \leq 2.0.
\end{equation}
For example, we can integrate them with a piecewise-constant function
\begin{eqnarray}
\label{eq:example_weight}
\omega(y) = \left\{\begin{array}{ll}
k,& y \geq 0,\\ 
-1,& y<0.\end{array}\right.
\end{eqnarray}
Here $k$ is some real number selected based on which specific factor we want to suppress. The resulting observable $\sigma^*$ would be a somewhat modified forward-backward scattering asymmetry:
\begin{eqnarray}
\sigma^* = \int dy \, \omega(y) \, \sigma_{YM}.
\end{eqnarray}


The factors $\sigma_{\mathrm{SM}}^*$, $\sigma_{a^2}^*$, $\sigma_{a v_l}^*$, and $\sigma_{a v_u}^*$ are linear functions of $k$:
\begin{eqnarray}
\label{eq:sigmastar_bounds}
\sigma^* & = & \sigma_\mathrm{SM}^* + a^2 \sigma_{a^2}^* + a v_u \sigma_{a v_u}^* + a v_l \sigma_{a v_l}^*, \nonumber\\
\sigma_{\mathrm{SM}}^* & = & (-23.7 + 26.8 \, k) \,\, \mathrm{pb} \pm (1.8 - 2.1 \, k) \,\, \mathrm{pb}, \nonumber\\ 
\sigma_{a^2}^* & = & (0.490 - 0.508 \, k) \,\, \mathrm{pb} \pm (0.032 - 0.033 \, k) \,\, \mathrm{pb}, \nonumber\\
\sigma_{a v_u}^* & = & (0.0147 - 0.0435 \, k) \,\, \mathrm{pb} \pm (0.0027 - 0.0025 \, k) \,\, \mathrm{pb}, \nonumber\\
\sigma_{a v_l}^* & = & (0.159 + 0.202 \, k) \, \mathrm{pb} \pm (0.014 + 0.017 \, k) \, \mathrm{pb}.
\end{eqnarray}

Let us construct an observable that is suitable for fitting of the axial-vector coupling $a$ and the  coupling to the up-quark vector current, $v_u$. That is, the factor $\sigma_{a v_l}^*$ has to be suppressed. We choose the suppression criteria
\begin{eqnarray}
\label{eq:k_criterium}
|\sigma_{a v_l}^*| < 0.01 |\sigma_{a^2}^*|, \qquad |\sigma_{a v_l}^*| < 0.01 |\sigma_{a v_u}^*|
\end{eqnarray}
to calculate $k$ in Eq. (\ref{eq:sigmastar_bounds}). By solving Eqs. (\ref{eq:k_criterium}) we obtain the resulting interval $-0.794 \, \leq \, k \, \leq \, -0.789$. If $k$ is set to $-0.79$ in Eq. (\ref{eq:sigmastar_bounds}), the resulting observable will contain only two unknown $Z'$ parameters:
\begin{eqnarray}
\sigma^* & = & \sigma_\mathrm{SM}^* + a^2 \sigma_{a^2}^* + a v_u \sigma_{a v_u}^*, \nonumber\\
\sigma_{\mathrm{SM}}^* & = & -44.8 \pm 3.5 \,\, \mathrm{pb}, \nonumber\\ 
\sigma_{a^2}^* & = & 0.890 \pm 0.058 \,\, \mathrm{pb}, \nonumber\\
\sigma_{a v_u}^* & = & 0.0480 \pm 0.0046 \,\, \mathrm{pb}.
\end{eqnarray}
This specific observable allows us to perform fitting of the $a$ and $v_u$ couplings.

There are two other possible observables in this approach: the one with suppressed $\sigma_{a v_u}^*$ and the one with suppressed $\sigma_{a^2}^*$. In Table \ref{tab:observables} we present the combinations of couplings that enter each of the proposed observables, together with the corresponding values of $k$ and $\sigma_{a^2}^*$, $\sigma_{a v_u}^*$, and $\sigma_{a v_l}^*$. Note, that we choose certain $k$ values from the corresponding intervals. Also, for the third variable the observable is close to a forward-backward asymmetry.

\begin{table*}[t]
\centering
\caption{\label{tab:observables}Couplings entering the observables, together with the corresponding values of $k$, the SM contribution $\sigma_{\mathrm{SM}}^*$, and the factors $\sigma_{a^2}^*$, $\sigma_{a v_u}^*$, and $\sigma_{a v_l}^*$}
\begin{tabular*}{\textwidth}{@{\extracolsep{\fill}}lccccc@{}}
\hline
couplings & $k$ & $\sigma_{\mathrm{SM}}^*$, pb & $\sigma_{a^2}^*$, pb & $\sigma_{a v_u}^*$, pb & $\sigma_{a v_l}^*$, pb \\
\hline
$a^2$, $a v_u$ & 
-0.79 & $-44.8 \pm 3.5$ & $0.892 \pm 0.058$ & $0.0490 \pm 0.0047$ & suppressed \\
$a^2$, $a v_l$ & 
0.35 & $-14.3 \pm 1.1$ & $0.312 \pm 0.021$ & suppressed & $0.230 \pm 0.020$ \\
$a v_u$, $a v_l$ & 
0.964 & $2.13 \pm 0.25$ & suppressed & $-0.0272 \pm 0.0003$ & $0.354 \pm 0.030$ \\
\hline
\end{tabular*}
\end{table*}

\section{\label{app:universality}Universal $Z'$ couplings to SM leptons}
We have mentioned earlier that the relations (\ref{RGrel1}) cover the $\chi$ model and {\it left-right} models. These models are widely considered in literature, in particular by the experimenters at the Tevatron and LHC. In these models $Z'$ is coupled to all three generations of fermions equally. Hence, it is useful to consider a special case of generation universality in our parameterization. The relations between couplings are extended as follows:
\begin{eqnarray}\label{RGrel1_univ}
&&
a_{q_d} = a_{l} = -a_{q_u} = -a_{\nu_l}=a,\nonumber\\
&&
v_{q_d} = v_{q_u} + 2 a, \qquad v_{l} = v_{\nu_l} + 2 a,\nonumber\\
&&
v_{q_u} = v_{q_c} = v_{q_t}, \qquad v_{e} = v_{\mu} = v_{\tau}.
\end{eqnarray}

This way the number of parameters in the Drell-Yan cross section is automatically reduced to only four:
\begin{eqnarray}
\label{eq:zpr_factors_univ}
\sigma_\mathrm{DY} &=& \sigma_\mathrm{SM} + \sigma_{Z'}, \nonumber\\
\sigma_{Z'} &=& a^2 \sigma_{a^2} + a v_l \sigma_{a v_l} 
+ a v_u \sigma_{a v_u} + 
v_u v_l \sigma_{v_u v_l}.
\end{eqnarray}

To obtain observables similar to the ones presented in the previous section, we can follow exactly the same procedure, but for the integration by $Y$. In that step no weight function is needed, since the contributions of the second generation do not introduce any new parameters in the cross section.

The whole integration scheme is reduced to only two steps:

1.~Integrate the cross section by $Y$ and $M$.

2.~Choose a coupling combination to suppress using the weight function $\omega(y)$ from Eq.~(\ref{eq:example_weight}).

Considering the same cuts, as in the general case, we obtain the observable: 
\begin{eqnarray}
\label{eq:sigmastar_bounds_univ}
\sigma^* & = & \sigma_\mathrm{SM}^* + a^2 \sigma_{a^2}^* + a v_u \sigma_{a v_u}^* + a v_l \sigma_{a v_l}^*, \nonumber\\
\sigma_{\mathrm{SM}}^* & = & (-102.2 + 116.6 \, k) \,\, \mathrm{pb} \pm (6.4 - 7.4 \, k) \,\, \mathrm{pb}, \nonumber\\ 
\sigma_{a^2}^* & = & (1.96 - 2.00 \, k) \,\, \mathrm{pb} \pm (0.11 - 0.11 \, k) \,\, \mathrm{pb}, \nonumber\\
\sigma_{a v_u}^* & = & (-0.0428 - 0.0247 \, k) \,\, \mathrm{pb} \nonumber\\
&\pm& (0.0101 - 0.0098 \, k) \,\, \mathrm{pb}, \nonumber\\
\sigma_{a v_l}^* & = & (0.729 + 0.909 \, k) \, \mathrm{pb} \nonumber\\
&\pm& (0.048 + 0.060 \, k) \, \mathrm{pb}.
\end{eqnarray}
Specific values of the parameter $k$ are presented in Table~\ref{tab:observables_univ}.

\begin{table*}[t]
\centering
\caption{\label{tab:observables_univ}Couplings entering the observables, together with the corresponding values of $k$, the SM contribution $\sigma_{\mathrm{SM}}^*$, and the factors $\sigma_{a^2}^*$, $\sigma_{a v_u}^*$, and $\sigma_{a v_l}^*$. The case of generation-universal couplings}
\begin{tabular*}{\textwidth}{@{\extracolsep{\fill}}lccccc@{}}
\hline
couplings & $k$ & $\sigma_{\mathrm{SM}}^*$, pb & $\sigma_{a^2}^*$, pb & $\sigma_{a v_u}^*$, pb & $\sigma_{a v_l}^*$, pb \\
\hline
$a^2$, $a v_u$ & 
-0.8015 & $-195.6 \pm 12.3$ & $3.56 \pm 0.20$ & suppressed & suppressed \\
$a^2$, $a v_l$ & 
-2.0 & $-335.3 \pm 21.2$ & $5.96 \pm 0.34$ & suppressed & $-1.09 \pm 0.07$ \\
$a v_u$, $a v_l$ & 
0.9805 & $12.12 \pm 0.82$ & suppressed & $-0.0670 \pm 0.0005$ & $1.62 \pm 0.11$ \\
\hline
\end{tabular*}
\end{table*}

Assumption of universality simplifies the proposed sche\-me not only by dropping a not-so-straightforward weighting function $W(M,Y)$, but also in terms of the quality of needed data: the same data that is used for measuring the forward-backward asymmetry \cite{Tevatron_AFB2014} can be used here.
Additionally, a one-parametric observable with a sign-definite parameter $a^2$ exists in this approach.

\section{\label{sec:discussion}Discussion}

The analysis of the LEP II data~\cite{GulovSkalozub:2009review,GulovSkalozub:2010ijmpa} resulted in obtaining upper bounds for $a^2$ and $v_l^2$ at 95\% CL, both of order of $0.1...1$ $m^2_{Z'}/\mathrm{TeV}^2$. In our recent paper \cite{GulovKozhushko:2014ijmpa} we have estimated these two couplings and $v_u^2$ to be $10^{-3}...10^{-1}$ $m^2_{Z'}/\mathrm{TeV}^2$ based on the results reported by ATLAS in 2012.
The results have been updated since then with the $\sqrt{S} = 8 \, \mathrm{TeV}$ \cite{CMS:2014zpr,ATLAS:2014zpr} data, but the
order of our estimates remains unchanged.

Let us assess the efficiency of the proposed observables based on these LEP- and LHC-driven estimates.
CDF collaboration presents the Drell-Yan cross section measurements \cite{CDF66} with a 1.5\%--2.5\% statistical and systematical uncertainties and with a 6\% luminosity uncertainty. As it is seen from Table \ref{tab:observables}, the obtained uncertainties for $\sigma_\mathrm{SM}^*$ are ~8\%--11\% (taking into account the considerations presented in \ref{app:kfactor}). It is safe to assume that this order of uncertainties will persist for the experimentally measured $\sigma_\mathrm{SM}^*$. It would be useful, if the proposed observables allowed to improve the bounds from~\cite{GulovSkalozub:2009review,GulovSkalozub:2010ijmpa,GulovKozhushko:2014ijmpa}. 
This requires the uncertainty calculated for the SM term has to be smaller than the value of the dominant $Z'$ contribution, e.g. $a^2 \sigma_{a^2}^*$ has to be over 3.5 pb for the first observable from Table~\ref{tab:observables}. In Table~\ref{tab:compare} we straightforwardly estimate efficiencies for all the proposed schemes, assuming all squared $Z'$ couplings to be of or\-der of $1$~$m^2_{Z'}/\mathrm{TeV}^2$, which corresponds to an optimistic LEP estimate, and considering a 2.6~TeV $Z'$ boson. It is also assumed, that $Z'$ couplings are positive and the $Z'$ effects are not suppressed by destructive interference between its contributions. One can see that the optimistic LEP-driven upper bounds will obviously be ruled out, as the magnitude of the $Z'$ contribution exceeds expected SM error. Since the considered energies are so far from the $Z'$ peak, the observable strongly depends on the magnitude of errors. In case if experimental data is precise enough, the observables might be of significant use for setting bounds on $Z'$ couplings. The leptonic universality case shows, that in principle hints of the $a^2$ coupling of order 0.1 $m_{Z'}^2/\mathrm{TeV}^2$ can be found at Tevatron, since the signal can be still over systematic errors.

If the estimation $0.1$ $m^2_{Z'}/\mathrm{TeV}^2$ obtained in \cite{GulovKozhushko:2014ijmpa} is considered, the new physics contributions from Table~\ref{tab:compare} are 10 times lower. In this case the bounds on the $Z'$ couplings can be set, if Tevatron data provides accuracy of order of $1$\%--$2$\% for the first two observables and $10$\% for the third observable.

\begin{table*}[t]
\centering
\caption{\label{tab:compare}Comparison of SM contribution uncertainty in $\sigma^*$ and estimated $Z'$ contributions, considering $m_{Z'} = 2.6 \, \mathrm{TeV}$ and all $Z'$ squared couplings $\sim 1$ $m^2_{Z'}/\mathrm{TeV}^2$. Only positive couplings are considered}
\begin{tabular*}{\linewidth}{@{\extracolsep{\fill}}lcccc@{}}
\hline
~&\multicolumn{2}{c}{Non-Universal}&\multicolumn{2}{c}{Universal}\\
\hline
couplings & $\delta\sigma^*_\mathrm{SM}/\sigma^*_\mathrm{SM}$ & $Z'$ part/$\sigma^*_\mathrm{SM}$ & $\delta\sigma^*_\mathrm{SM}/\sigma^*_\mathrm{SM}$ & $Z'$ part/$\sigma^*_\mathrm{SM}$ \\
\hline
$a^2$, $a v_u$ & $7.7\%$ & $-14.1\% \pm 0.8$\% & $6.3\%$ & $-12.2 \pm 0.6$ \\
$a^2$, $a v_l$ & $7.4\%$ & $-25.5\% \pm 1.8$\% & $6.3\%$ & $-9.8\% \pm 0.5\%$ \\
$a v_u$, $a v_l$ & $11.6\%$ & $103.7\% \pm 9.7$\% & $6.8\%$ & $86.7\% \pm 6.1\%$ \\
\hline
\end{tabular*}
\end{table*}

Another way to check if the proposed observable is useful is to compare them to the current coupling values for SM $Z$ boson measured at the LEP, Tevatron, HERA and LHC. Those can be found on p. 27 of \cite{rpp}. For example, let us consider $Z$ couplings to the $u$ quark.
By comparing the utilized parameterization (\ref{ZZplagr}) with the SM $Z$ couplings \cite{rpp}, we obtain the following relations:
\begin{eqnarray}
a v_u &=& \frac{1}{2} \left( \frac{\sqrt{4 \pi \alpha}}{\cos \theta_W \sin \theta_W} \frac{m_{Z'}}{m_Z} \right)^2 \left(\frac{1}{2} - g_V^u - \frac{4}{3} \sin \theta_W^2 \right), \nonumber\\
a^2 &=& \frac{1}{2} \left( \frac{\sqrt{4 \pi \alpha}}{\cos \theta_W \sin \theta_W} \frac{m_{Z'}}{m_Z} \right)^2 \left(g_A^u - \frac{1}{2} \right).
\end{eqnarray}
Here we have also substituted the $Z$--$Z'$ mixing angle from Eq. (\ref{RGrel2}). The SM $Zu\bar{u}$ couplings are \cite{rpp} $g_V^u = 0.25^{+0.07}_{-0.06}$ and $g_A^u = 0.50^{+0.04}_{-0.06}$. This leads to the bounds on the $Z'$ couplings:
\begin{eqnarray}
-0.35 \leq a v_u \frac{\mathrm{TeV}^2}{m_{Z'}^2} \leq 3.77, \quad
0 \leq a^2 \frac{\mathrm{TeV}^2}{m_{Z'}^2} \leq 1.88.
\end{eqnarray}
These bounds are of the same order as the ones obtained from LEP data analysis in \cite{GulovSkalozub:2009review,GulovSkalozub:2010ijmpa}. Therefore, we expect to improve these bounds approximately by one order of magnitude to about $0.1$ TeV$^2/m^2_{Z'}$.

An observable similar to the one proposed here can be constructed for the LHC case, although some modifications are required to suppress contributions from the second generation of fermions. Higher luminosity should lead to a significantly higher precision for the $Z'$ couplings.

Neither LEP data nor Tevatron or LHC data show any explicit indications of the Abelian $Z'$. This provides motivation to investigate models with the so called leptophobic $Z'$ \cite{leptophobic:1,leptophobic:2,leptophobic:3,leptophobic:4,leptophobic:5,leptophobic:6,leptophobic:7}. In these models $Z'$ boson couplings to the SM leptons are strongly suppressed compared to the quark couplings. Among other things, this parameterization allows to explain deviations of the precision electroweak data from the SM by introducing $Z'$ with the mass close to $m_Z$ \cite{Dermisek2011}. From the Lagrangian in Eq. (\ref{ZZplagr}) and the relations in Eq. (\ref{RGrel1}) it follows that in the leptophobic case $v_l$, $a_l$, and $a_q$ are small compared to $v_q$, and the leading $Z'$ contributions to the cross section are
\begin{eqnarray}
\label{eq:zpr_factors_leptophobic}
\sigma_\mathrm{DY} &=& \sigma_\mathrm{SM} + \sigma_{Z'}, \nonumber\\
\sigma_{Z'} &=& 
a v_u \sigma_{a v_u} + 
v_u v_l \sigma_{v_u v_l} 
\nonumber\\
&& 
+ a v_c \sigma_{a v_c} + 
v_c v_l \sigma_{v_c v_l} + O(a^2,a v_l).
\end{eqnarray}
After applying all the integrations discussed in Section~\ref{sec:theobservable}, we end up with the observable where only the term $a v_u \sigma^*_{a v_u}$ survives. This observable is one-para\-me\-tric: 
\begin{eqnarray}
\label{eq:sigmastar_leptophobic}
\sigma^* & = & \sigma_\mathrm{SM}^* + a v_u \sigma_{a v_u}^*.
\end{eqnarray}
The numerical values are the same as in the second line of Table \ref{tab:observables}.

Our results obtained for the dimuon cuts can be easily recalculated for dielectrons, taking into account the difference between event selection criteria for muons and electrons. For example, the CDF collaboration selects muons with maximum pseudorapidity $|\eta_\mu| = 1.0$ \cite{CDF_rapidity_mumu_2011}. For electrons this value is $|\eta_{e}| = 2.8$ \cite{CDF_rapidity_ee_2010}, therefore, the value of $Y_m$ for the $p\bar{p} \to e^+ e^-$ process is higher than for the dimuon case. This leads to different weight functions and values of $k$.


\section{\label{sec:conclusions}Conclusions}

The data analysis performed by the LHC and Tevatron collaborations resulted in setting model-dependent lower bounds on the $Z'$ mass. In that analysis only the high-energy range of the Drell-Yan cross section was considered. In our paper we propose different approach that allows to search for a $Z'$ signal in the $p\bar{p} \to l^+ l^-$ process at the energies near $m_Z$.
In this range the most important contributions at the $Z$ peak come from the $Z-Z'$ mixing.
The approach utilizes the renormalization-group relations between the effective $Z'$ couplings. Therefore, in case no signal is observed one would still be able to derive constraints for different $Z'$ models and compare them to the ones presented in \cite{CMS:2014zpr,ATLAS:2014zpr}.

The proposed prescription includes the following steps:

1.~The triple-differential cross section of the Drell-Yan process is expressed in terms of three kinematic variables: the mass of an intermediate state, $M$, the intermediate-state rapidity $Y$, and the relative rapidity of a lepton pair, $y$. This cross section contains six unknown combinations of the $Z'$ couplings: $a^2$, $a v_u$, $a v_l$, $a v_c$, $v_u v_l$, and $v_c v_l$.

2.~The cross section is integrated by $Y$ over the symmetric range $[-Y_\mathrm{max}; Y_\mathrm{max}]$ with the weight function $W(M,Y)$ defined in Eq.~(\ref{eq:observ_Y_wf}).
The integration limits have to be determined for specific final state ($e^+ e^-$ or $\mu^+ \mu^-$) and detector individually. The function $W(M,Y)$ has to be adjusted in such way, that the PDF factors for the second-generation quarks, $F_{c\bar{c}}(M)$ and $F_{s\bar{s}}(M)$, amount to less than 1\% of the PDF factor $F_{u\bar{u}}(M)$. In the dimuon case this is a linear function of $M/m_Z$. As a result we exclude $a v_c$ and $v_c v_l$ from the cross section.

3.~Integrate the cross section by $M$ over the $Z$ boson peak range: either $66 \mathrm{~GeV} \leq M \leq 116 \mathrm{~GeV}$ or $71 \mathrm{~GeV} \leq M \leq 111 \mathrm{~GeV}$, or any other range with bounds symmetric with respect to $m_Z$. These bounds have to be large enough, so that one could neglect the masses of the $u$, $d$, $c$, and $s$ quarks compared to $M$.
This integration suppresses the contribution of $v_u v_l$ to the cross section.

4.~The integration by $y$ with the properly adjusted weight function $\omega(y)$ from Eq.~(\ref{eq:example_weight}) allows to suppress either $a^2$, $a v_u$, or $a v_l$.

We provided the example of how to use the proposed procedure. The numerical values of the cutoffs $Y_\mathrm{mid}$, $Y_\mathrm{max}$ and $k$ may vary, as they depend on specific experimental conditions, e.g. bin structure and available data. For example, the $Y_m$ value can easily be moved closer to the detector coverage limit, and the weighting functions $W(M,Y)$ and $\omega(y)$ will have to be adjusted accordingly, but the general scheme, including the qualitative form of the weighting functions, will remain unchanged.

The obtained observables can be used in fitting the experimental data on the $p\bar{p} \to l^+ l^-$ scattering collected by the Tevatron collaborations. This allows to constrain the $Z'$ vector axial-vector couplings to SM fermions.

In case of the leptophobic $Z'$ boson, there is a one-parametric observable constraining the combination of couplings $a v_u$. Also, a one-parametric observable exists in the parameterization with universal interactions of $Z'$ with lepton generation. This observable allows to constrain a sign-definite coupling $a^2$.

There is a large amount of data on leptonic scattering processes collected in the LEP and LEP II experiments. The second observable in Table \ref{tab:observables} contains the coupling combinations $a^2$ and $a v_e$ that also enter lepton scattering processes. It seems to be useful for combined fits of the LEP and Tevatron data.

\section{\label{sec:acknowledgements}Acknowledgements}

This research has been partially supported by funds of the Abdus Salam ICTP under the OEA Program (AC-88).

\appendix

\section{\label{app:kfactor}Generating the K-factor}

\begin{figure}[t]
\centering{
\resizebox{1\linewidth}{!}{
\includegraphics{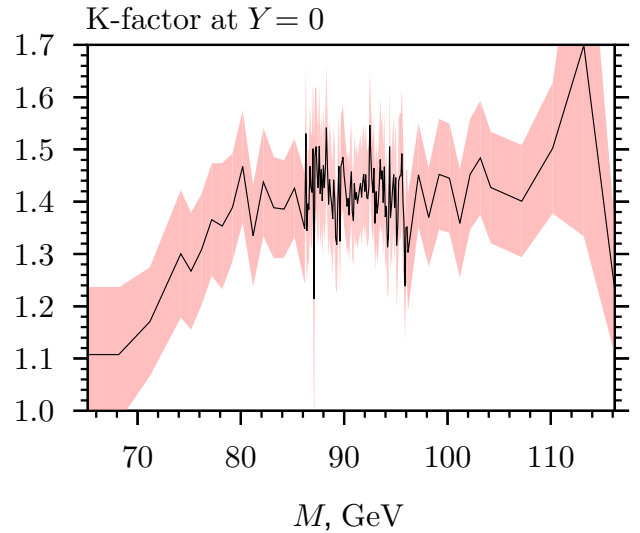}}}
\caption{\label{fig:kfactornofit}The K-factor values at $Y=0$ obtained with \textit{FEWZ}. The numerical uncertainty and the 90\% CL PDF error are included. The $M$ bin sizes used for computation are 3~GeV for $M = 65~\mathrm{GeV}..75~\mathrm{GeV}$ and $M = 107~\mathrm{GeV}..116~\mathrm{GeV}$, 1~GeV for $M = 76~\mathrm{GeV}..85~\mathrm{GeV}$ and $M = 97~\mathrm{GeV}..104~\mathrm{GeV}$, and 0.1~GeV $M = 86~\mathrm{GeV}..96~\mathrm{GeV}$. Such binning was chosen in attempt to obtain correct $Z$ peak form and save computation time}
\end{figure}

As it was mentioned in Section \ref{sec:ZprDY}, the calculation of K-factor for the double-differential cross section requires significant computational resources. At the time of paper preparation we had only an 8-core i7 processor available, hence the data obtained using \textit{FEWZ 2.1.1} contained significant numerical error. The generated data along with input parameters are available in the file storage \cite{KfactorFiles}. For our calculations we needed an estimation of the K-factor, which could be included into the cross section and provide a realistic estimation of QCD corrections. As it is seen from Figure \ref{fig:kfactornofit}, the numerical uncertainty leads to a non-smooth K-factor. To calculate some smooth estimate suitable for calculations, we fit the obtained computational data with a function of $M$ and $Y$ using a $\chi^2$ method.

The fitting polynomial is chosen as
\begin{eqnarray}
\label{eq:ktheor}
K_{\mathrm{theor}} = p_0 + p_1 M + p_2 Y^2.
\end{eqnarray}

The $\chi^2$ function is computed as a sum over all bins:
\begin{eqnarray}
\label{eq:chi2}
\chi^2 = \sum \frac{(K_\mathrm{comp}^\mathrm{mean} - K_\mathrm{theor})^2}{\delta K_\mathrm{comp}^2}.
\end{eqnarray}
Here $K_\mathrm{comp}^\mathrm{mean}$ and $\delta K_\mathrm{comp}^2$ denote the mean value and error (numerical and PDF) computed with \textit{FEWZ}. For 3 degrees of freedom and 95\% CL the $\chi^2$ percentile is $7.81$. The minimum $\chi^2$ value and the corresponding polynomial coefficients are presented in Table \ref{tab:fitkfactor}.

\begin{table}[t]
\centering
\caption{\label{tab:fitkfactor}The minimum value of $\chi^2$ along with fitted coefficients from Eq. (\ref{eq:ktheor}).}
\begin{tabular*}{0.5\linewidth}{@{\extracolsep{\fill}}ll@{}}
\hline
$\chi^2_{\mathrm{min}}/\mathrm{bin}$ & $1.13$ \\
$p_0$ & $1.30$\\
$p_1$ & $0.00145$\\
$p_2$ & $0.00735$\\
\hline
\end{tabular*}
\end{table}

\begin{figure}[t]
\resizebox{1\linewidth}{!}{
\includegraphics{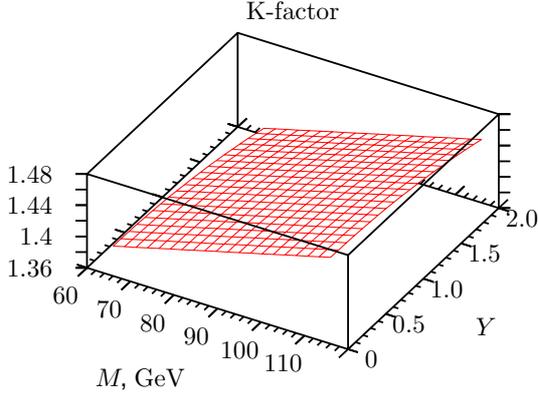}}
\caption{\label{fig:kfactorfit}The approximate mean value of the K-factor}
\end{figure}

The mean value for K-factor is plotted in fig. \ref{fig:kfactorfit}. The 95\% CL bounds for the K-factor are of order of 1\%, so we do not include them into our consideration.

The obtained estimate is rather rough. Nevertheless, it corresponds to the NNLO estimates used in \cite{2004Carena}. Another point is that the soft NLO corrections are expressed by the relation
\begin{eqnarray}
\label{eq:NLO}
K_{\mathrm{NLO}} = 1 + \frac{4}{3} \frac{\alpha_S(Q^2)^\mathrm{NLO}}{2 \pi} \left(1 + \frac{4}{3} \pi^2\right),
\end{eqnarray}
where $\alpha_S(Q^2)^\mathrm{NLO}$ varies from 0.135 to 0.123 for $Q$ from 66 GeV to 116 GeV. Hence, our computed K-factor is consistent with this NLO prediction of 1.37--1.40. These considerations allow us to use the fitted function in our estimation.

\section{Partonic cross sections\label{app:cs}}

In this section we provide expressions for partonic differential cross sections $u\bar{u} \to l^+ l^-$ and $d\bar{d} \to l^+ l^-$ in Eq. (\ref{eq:MYy_cs}) with $Z'$ contributions in the parameterization (\ref{grgav}).

Each partonic cross section is written as in Eq. (\ref{eq:partcs}). The $\hat{\sigma}$ factors have the following general form similar to Eq. (\ref{eq:sigma1_yM}):
\begin{eqnarray}
\label{eq:sigma1_yM}
\hat{\sigma} = \frac{\cosh 2y}{\cosh^4 y} \left[ A(M) \tanh 2y + S(M) \right],
\end{eqnarray}
The $M$ dependence of even part $S(M)$ and odd part $A(M)$ is expressed in terms of the resonant functions provided by Eqs. (\ref{eq:RF}).

In these expressions we use the following notations:
\begin{eqnarray}
\zeta &=& M/m_z, \quad \xi = m_Z/m_{Z'}, \nonumber\\
s_W &=& \sin \theta_W, \quad c_W = \cos \theta_W, \nonumber\\
p_1 &=& 16 s_W^2 c_W^2, \quad p_2 = 3-10s_W^2, \nonumber\\
p_3 &=& 5 - 8 s_W^2, \quad p_4 = 3 - 8 s_W^2, \nonumber\\
p_6 &=& 4 s_W^2 -1, \quad p_7^\pm = 3 \pm 4 s_W^2, \nonumber\\
p_8 &=& 8 s_W^2 + 1.
\end{eqnarray}

The contributions to the $u\bar{u}$ cross section are as follows. For $\hat{\sigma}_\mathrm{SM}$:
\begin{eqnarray}
A^{u\bar{u}}_\mathrm{SM} &=& \frac{2\alpha^2 \pi f_1 \zeta^2}{9 p_1^2 m_Z^2} \left(p_1 \frac{\zeta^2-1}{\zeta^2} - p_4 p_6\right), \nonumber\\
S^{u\bar{u}}_\mathrm{SM} &=& 
\frac{2 \alpha^2 \pi}{27 m_Z^2} \left[ \frac{1}{\zeta^2}\right. \nonumber\\
&+& \left. \zeta^2 f_1 \left(\frac{p_2^2 + 4 s_W^2 p_3^2}{p_1^2} - \frac{p_4 p_6}{p_1} \frac{\zeta^2-1}{\zeta^2} \right)\right]. \nonumber
\end{eqnarray}

For $\hat{\sigma}_{a^2}$:
\begin{eqnarray}
A^{u\bar{u}}_{a^2} &=& 
\frac{\alpha \xi^2}{432 p_1^2 m_Z^2} \left[6 p_1 \left\lbrace f_2'(1+\xi^2)^2 - 2 f_2\right\rbrace \right. \nonumber\\
&+& \left. 3 p_4 p_6 \left\lbrace 4 \zeta^2 f_1 - f_3 \left(1+4\xi^2(1+\xi^2) \right) \right\rbrace\right], \nonumber\\
S^{u\bar{u}}_{a^2} &=& 
\frac{\alpha \xi^2}{432 p_1^2 m_Z^2} \nonumber\\
&\times& \left[ -4\zeta^2 f_1 (p_2^2 + 6 p_3 s_W^2) -2 p_1 p_4 p_6 \xi^4 f'_2\right. + f_3  \nonumber\\
&\times& \left. \left\lbrace 9 + 4 \xi^2 \left( p_2^2 + 6 p_3 s_W^2 + \xi^2 ( p_2^2 + 4 s_W^4 p_3^2 ) \right) \right\rbrace \right]. \nonumber
\end{eqnarray}

For $\hat{\sigma}_{av_u}$:
\begin{eqnarray}
A^{u\bar{u}}_{av_u} &=& 
\frac{\alpha \xi^2}{48 p_1^2 m_Z^2} p_6 \left( 2 \zeta^2 f_1 - f_3 (1+2 \xi^2) \right), \nonumber\\
S^{u\bar{u}}_{av_u} &=& 
\frac{\alpha \xi^2}{144 p_1^2 m_Z^2} \nonumber\\ 
&\times& \left[ 2 \zeta^2 f_1 p_4 (p_6 - 8 s_W^4) + 2 p_1 p_6 (f_2 - \xi^2 f'_2) \right.
\nonumber\\
&+& \left. f_3 \left( p_2 + 2 \xi^2 (32 s_W^4 p_3 - p_4 p_6) \right) \right]. \nonumber
\end{eqnarray}

For $\hat{\sigma}_{av_e}$:
\begin{eqnarray}
A^{u\bar{u}}_{av_e} &=& 
\frac{\alpha \xi^2}{144 p_1^2 m_Z^2} p_4 \left( 2 \zeta^2 f_1 - f_3 (1+2 \xi^2) \right), \nonumber\\
S^{u\bar{u}}_{av_e} &=& 
\frac{\alpha \xi^2}{432 p_1^2 m_Z^2} \nonumber\\ 
&\times& \left[ 2 p_1 p_4 (f_2 - \xi^2 f'_2) - 3 \zeta^2 f_1 p_6 (3 p_4 + 32 s_W^4) \right.
\nonumber\\
&+& \left. f_3 p_6 \left( 9 + 2 \xi^2 (3 p_4 + 32 s_W^4) \right) \right]. \nonumber
\end{eqnarray}

For $\hat{\sigma}_{v_u v_e}$:
\begin{eqnarray}
A^{u\bar{u}}_{v_u v_e} &=& 
- \frac{\alpha \xi^2}{48 p_1^2 m_Z^2} f_3, \nonumber\\
S^{u\bar{u}}_{v_u v_e} &=& 
\frac{\alpha \xi^2}{144 p_1^2 m_Z^2} \left( 2p_1 f'_2 - p_4 p_6 f_3 \right). \nonumber
\end{eqnarray}

The contributions to the $d\bar{d}$ cross section are as follows. For $\hat{\sigma}_\mathrm{SM}$:
\begin{eqnarray}
A^{d\bar{d}}_\mathrm{SM} &=& \frac{2\alpha^2 \pi f_1 \mu^2}{9 p_1^2 m_Z^2} \left(p_2 \frac{\zeta^2-1}{2\zeta^2} - p_7^- p_6\right), \nonumber\\
S^{d\bar{d}}_\mathrm{SM} &=& 
\frac{\alpha^2 \pi}{27 m_Z^2} \left[ \frac{1}{2\zeta^2}\right. \nonumber\\
&+& \left. \zeta^2 f_1 \left(\frac{4 p_4^2 + p_1^2}{2p_1^2} - \frac{p_7^- p_6}{p_1} \frac{\zeta^2-1}{\zeta^2} \right)\right]. \nonumber
\end{eqnarray}

For $\hat{\sigma}_{a^2}$:
\begin{eqnarray}
A^{d\bar{d}}_{a^2} &=& 
\frac{\alpha \xi^2}{144 p_1^2 m_Z^2} \left[p_1 \left\lbrace f_2'(1+\xi^2)^2 - 12 f_2\right\rbrace \right. \nonumber\\
&+& \left. p_6 \left\lbrace f_3 (1+2\xi^2) \left(8 s_W^2 - p_7^- (2\xi^2-1) \right) - 16 \zeta^2 f_1 s_W^2\right\rbrace\right], \nonumber\\
S^{d\bar{d}}_{a^2} &=& 
\frac{\alpha \xi^2}{432 p_1^2 m_Z^2} \left[ 2 \xi^2 f'_2 p_1 (3 p_6 + \xi^2 p_8)\right. \nonumber\\
&+& f_1 \left( p_1 p_6 (1-\zeta^2) - 32 \zeta^2 s_W^2 (p_2 + 12 s_W^4) \right) \nonumber\\
&+& f_3 \left. \left\lbrace p_4 (4 p_4 \xi^4 - 3) + 32 s_W^2 \xi^2 (p_2 + 12 s_W^4) + \xi^4 p_1^2 \right\rbrace \right]. \nonumber
\end{eqnarray}

For $\hat{\sigma}_{av_u}$:
\begin{eqnarray}
A^{d\bar{d}}_{av_u} &=& 
\frac{\alpha \xi^2}{48 p_1^2 m_Z^2} p_6 \left( f_3 (1+2 \xi^2) - 2 \zeta^2 f_1 \right), \nonumber\\
S^{d\bar{d}}_{av_u} &=& 
\frac{\alpha \xi^2}{144 p_1^2 m_Z^2} \left[ 2 \zeta^2 f_1 p_7^- (6 s_W^4 - p_6) \right.
\nonumber\\ 
&+& \left. p_1 p_6 (\xi^2 f'_2 - f_2) - f_3 p_7^- \left( 1 + 2 \xi^2 (6 s_W^4 - p_6) \right) \right]. \nonumber
\end{eqnarray}

For $\hat{\sigma}_{av_e}$:
\begin{eqnarray}
A^{d\bar{d}}_{av_e} &=& 
\frac{\alpha \xi^2}{144 p_1^2 m_Z^2} \left( 2 \zeta^2 f_1 p_7^- + f_3 (p_7^+ - \xi^2 p_7^-) \right), \nonumber\\
S^{d\bar{d}}_{av_e} &=& 
\frac{\alpha \xi^2}{432 p_1^2 m_Z^2} \left[ p_1 p_7^- (f_2 - \xi^2 f'_2) + 6 p_1 f'_2 \right. \nonumber\\
&-& 2 \zeta^2 p_6 f_1 (3 p_7^- + 8 s_W^4)
\nonumber\\
&+& \left. f_3 p_6 \left( 2 \xi^2 (3 p_7^- + 8 s_W^4) - 3 p_4 \right) \right]. \nonumber
\end{eqnarray}

For $\hat{\sigma}_{v_u v_e}$:
\begin{eqnarray}
A^{d\bar{d}}_{v_u v_e} &=& \frac{\alpha \xi^2}{48 p_1^2 m_Z^2} f_3, \nonumber\\
S^{d\bar{d}}_{v_u v_e} &=& 
\frac{\alpha \xi^2}{144 p_1^2 m_Z^2} \left( p_1 f'_2 - p_7^- p_6 f_3 \right). \nonumber
\end{eqnarray}

Same partonic cross section are provided in the data set \cite{factorsandmodelfile}, which is composed in a form suitable for computational packages.

\end{document}